# Telescope Spectrophotometric and Absolute Flux Calibration, and National Security Applications, Using a Tunable Laser on a Satellite

J. Albert[1], W. Burgett[2], J. Rhodes[3,4]

May 19, 2006

## 1. Overview

Absolute photon flux uncertainty, and relative uncertainties in atmospheric transmission as a function of wavelength, limit photometric precision in ground-based telescopes. As illustrated in Figs. 1-3, improvement to significantly better than 1% is necessary to prevent photometric precision from being a limiting factor in upcoming cosmological measurements using type Ia supernovae [1] as well as weak gravitational lensing [2]. The most precise presently available standard for flux calibration is Vega, which is known to uncertainties of approximately 2% in the optical and 4% in the near infrared [3,4]. Although those factors are *absolute* flux uncertainties, whereas *relative* flux uncertainties as a function of wavelength are what are most critical for precise redshift measurements, it is necessary to have excellent calibration for both relative and absolute flux in order for both precise and accurate comparisons to standard luminosity sources to be made.

Recently Stubbs *et al.* have demonstrated the use of a tunable laser for relative photometric flux calibration [5]. A tunable laser can provide a precise amount of monochromatic light in a range of frequencies. A NIST-calibrated photodiode in parallel with the focal plane provides the absolute integrated radiometric scale. This technique has several benefits: it provides a complete characterization of instrumental response as a function of wavelength with a relatively simple and convenient system. However, a ground-based tunable laser cannot provide a calibration of the transmission loss and photometric shifts induced by the atmosphere. Such an extrapolation requires atmospheric model dependence and/or use of an absolute stellar source such as Vega, once again limiting the photometric uncertainty to 2-4%.

The presence of a standard "star" in space with flux known to significantly better than 1% would permit a greatly improved direct calibration of photometric uncertainty. In principle, perfect knowledge of only a single absolute spectrum from a source in space (or spectra from multiple sources) in conjunction with a tunable laser source on the ground would be sufficient, under the assumption that one can fully and properly model any potential ambiguous intra-spectral shifts due to absorption and re-emission in the atmosphere. Having a tunable monochromatic source in space would further reduce the dependence on models of atmospheric absorption and re-emission.

---
[1] Department of Physics, California Institute of Technology, MS356-48, Pasadena, CA 91125, U.S.A.; justin@hep.caltech.edu
[2] Institute for Astronomy, University of Hawaii, 2680 Woodlawn Dr., Honolulu, HI 96822-1867, U.S.A.
[3] Jet Propulsion Laboratory, Pasadena, CA 91109, U.S.A.
[4] Department of Astronomy, California Institute of Technology, MS105-24, Pasadena, CA 91125, U.S.A.

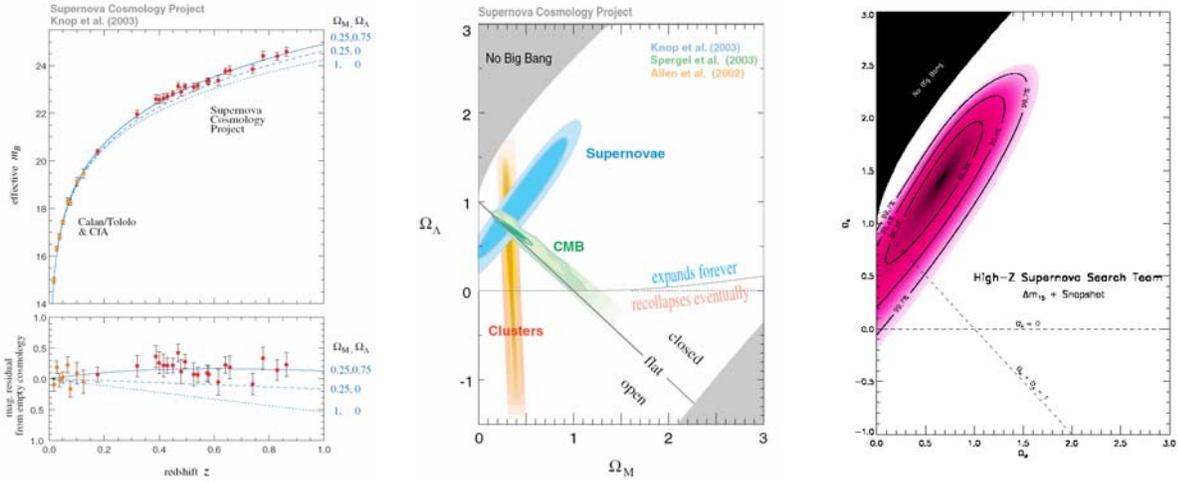

*Figure 1*: Constraints on the $\Omega_M$-$\Omega_\Lambda$ plane from type Ia supernova measurements from the Supernova Cosmology Project [6] (left two plots) and the High-z Supernova Search [7] (right plot). As the leftmost plot shows, a relative error of O(1%) in redshift measurements results in O(.03) errors on $\Omega_M$ and $\Omega_\Lambda$. Thus relative errors of O(1%) in flux as a function of wavelength can potentially result in up to O(.03) errors on fundamental cosmological parameters $\Omega_M$ and $\Omega_\Lambda$. The equation of state parameters $w_0$ and $w_a$ are even more sensitive.

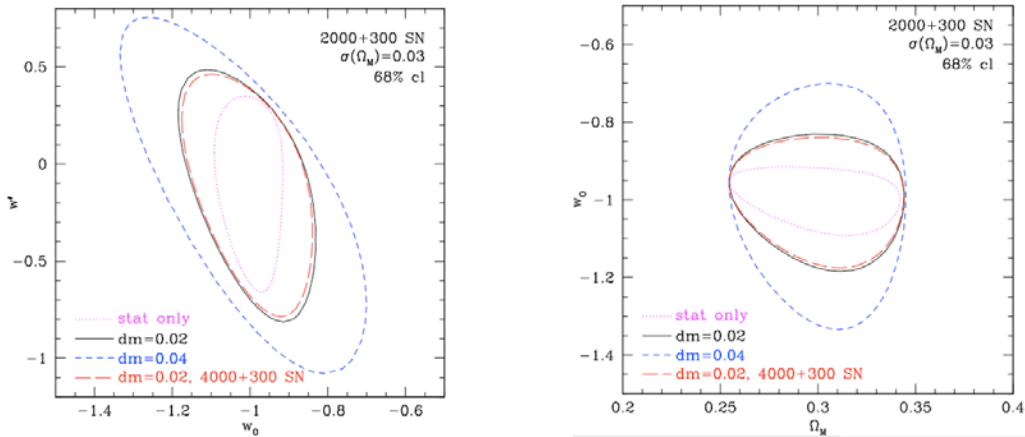

*Figure 2*: From Kim, Linder, Miquel, and Mostek [1]: Dark energy parameter contours $w'$ vs. $w_0$ (left) and $w_0$ vs. $\Omega_M$ (right) from type Ia supernovae for three irreducible systematic cases. The central reference contour has only the intrinsic statistical error. The parameter error increases substantially as the systematic is increased from $dm = 0.02$ to $0.04$. The two closely-paired contours show that doubling the number of supernovae does not significantly decrease parameter uncertainty at this point; the cosmological parameter uncertainty is primarily dependent on maintaining low systematic uncertainty. In the right plot, the $\Omega_M$ parameter error stays constant since the a priori constraint of $\sigma(\Omega_M) = 0.03$ dominates the estimation.

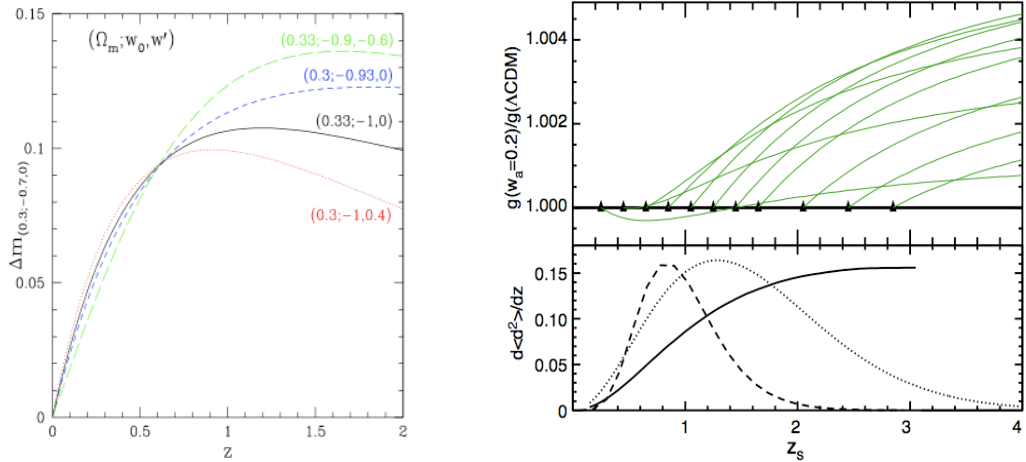

*Figure 3*: (Left) From Huterer and Linder [8]: type Ia supernova differential magnitude vs. redshift for different values of cosmological parameters ($\Omega_M, w_0, w'$). The difference between models is less than 0.05 magnitudes, requiring 2% flux calibration uncertainty to clearly separate. (Right) From Bernstein and Jain [2]: cosmological information from weak lensing surveys with source galaxy redshift information included. "The top panel shows the fractional change in the geometric factor $g_{\ell s} = (\chi_s - \chi_\ell)/\chi_s$ when we shift from a pure $\Lambda$CDM universe to one with $w_a = 0.2$; this should be discernable at $1\sigma$ significance in the SNAP survey. The horizontal axis is at $z_s$ and each line corresponds to a different $z_\ell$. The triangle at the end of each line marks $z_\ell$…. The cosmological information is carried in the departures of each line from horizontal; these departures are small, amounting to only a few parts in $10^{-3}$ change in the induced background distortion. The smallness of this signal implies that the calibration of the distortion measurement and the source redshifts must be accurate to roughly a part in $10^3$. The lower panel plots the assumed source redshift distribution (dotted line) and the expected source variance per unit redshift (solid line) using estimated non-linear power spectra. The dashed line shows the relative contribution of different lens planes to the constraint on $w_a$."

Aside from fundamental astrophysical applications, a laser in space also potentially has a panoply of defense and national security benefits; these have been known for quite some time [9]. A laser in space could provide extensive and highly tactically beneficial ground target illumination and space communication capabilities that are presently unavailable to U.S. forces.[5]

This paper provides a brief overview of the astrophysical and military capabilities of a tunable laser in space that is constructed entirely with presently available technology, and capable of providing a permanent nightly 0.1% calibration in a broad range of frequencies for all telescopes in Hawaii and Chile, as well as providing new tactical and national security capabilities for U.S. forces. We provide a list of specifications of the components of such a satellite (the laser, the orbit, pointing and slew requirements, telemetry, etc.) to achieve these required astrophysical and military goals.

---

[5] Note that the possibility of an extremely powerful laser in space would potentially provide the ability to destroy ballistic missiles in-flight, as was studied extensively in the SDI program.

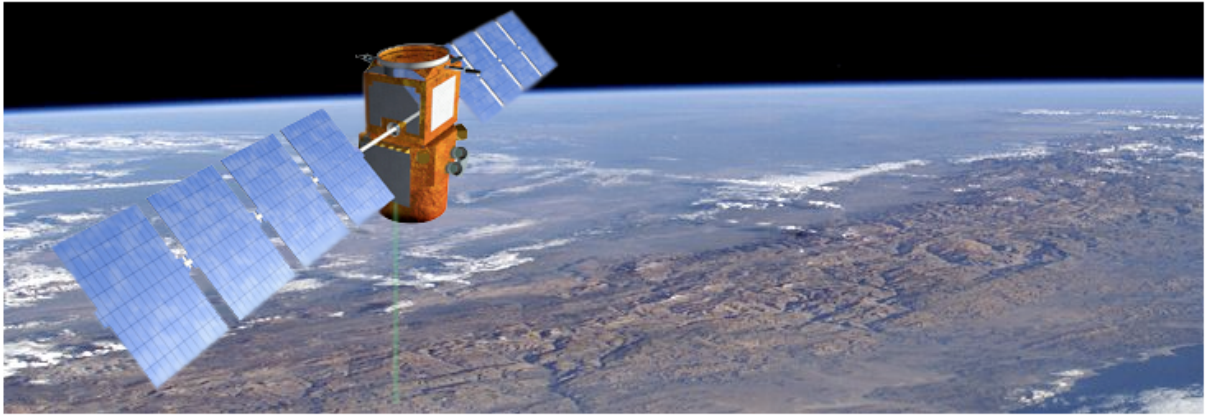

*Figure 4*: Artistic impression of the CALIPSO (Cloud-Aerosol Lidar and Infrared Pathfinder) Satellite, for atmospheric observations [10]. CALIPSO was launched on April 28, 2006, and is presently in a polar orbit, in formation with five other atmospheric satellites, at 705 km altitude. CALIPSO contains a Nd:YAG laser, producing light at 523 nm and 1064 nm wavelengths, which can accurately point at a given location on the globe. We plan to take test data at Palomar Observatory later this year with light pulses from the CALIPSO laser.

Lasers in space also have the potential to dramatically improve satellite communication. Inter-satellite communication via laser offers virtually unlimited bandwidth, at low power, in an unregulated portion of the electromagnetic spectrum. Two-way Earth-satellite communication has recently been demonstrated by the Messenger spacecraft across 24 million kilometers of space to a ground station at Goddard [11]. Inter-satellite communication has also first been demonstrated this past year between the Kirary and Artemis satellites (a Japanese satellite in low-Earth orbit and a geosynchronous European satellite respectively) [12]. This nascent technology has been eagerly anticipated for over 40 years (since the initial patent on the laser [13]) due to the continued great promise in increased bandwidth and reduced power consumption, and is now a clear path for development for use in future interplanetary exploration, scientific, communication, and military satellite systems.

In addition, atmospheric science is also a large motivation for space-based lasers. The CALIPSO satellite, launched by NASA in April of this year (see Fig. 4), contains a Nd:YAD LIDAR laser for cloud formation observations, producing wavelengths of 532 nm and 1064 nm. We in fact intend to take test data at Palomar Observatory using light pulses from the CALIPSO laser. A tunable laser, in addition to astrophysical and military benefits, would allow atmospheric scientists to explore a far greater range of wavelengths than with the present (groundbreaking) CALIPSO satellite [10].

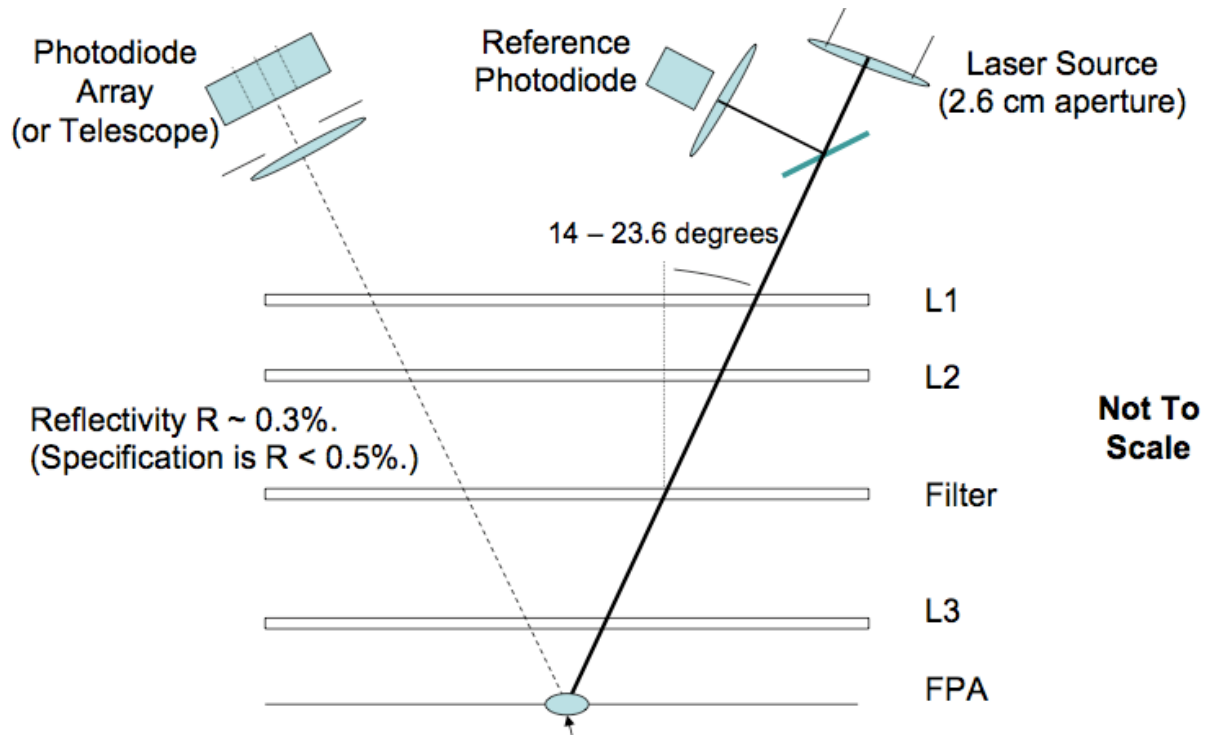

*Figure 5*: From D. Burke [14]: cartoon of ground-based tunable laser throughput and photometry calibration (Stubbs, Burke *et al.*). The NIST-calibrated reference photodiode provides an overall absolute calibration of the laser flux at the source. L1-3 are refractive optics specific to the LSST camera. Putting the right-hand side of the setup in space reduces dependence on accuracy of atmospheric model, and can correct different points in the sky (as well as providing a precision calibration of stars passing nearby the path of the satellite, which can then be used as very well-calibrated sources themselves).

The wide rage of fundamental applications of a space-based tunable laser in astrophysical, military, and atmospheric science provide a major motivation for a small, cost-effective mission.

## 2. Telescope spectrophotometry and absolute flux calibration

Figure 5 shows a tentative calibration diagram for the LSST telescope camera (as one example) using a tunable laser and NIST-calibrated photodiode. Such a calibration can fully characterize the instrumental response of the telescope, but does not calibrate the additional flux uncertainties due to the atmosphere.

Consider a satellite in medium Earth orbit at 20000 km (a similar altitude to the present GPS satellites [15]). Such a satellite travels at twice the Earth's rotation speed – ~12000 km/hr relative to the Earth's surface, requiring a maximum slewing rate of 0.6°/min to follow from the ground (see Fig. 6), within the capability of all ground telescopes, including of course LSST or Pan-STARRS, since this is the same relative rotation speed to Earth, as Earth has with respect to the night sky. Medium-Earth orbits such as those used by GPS satellites can also potentially pass over both Hawaii and Chile, allowing calibration of major observatories in both locations.

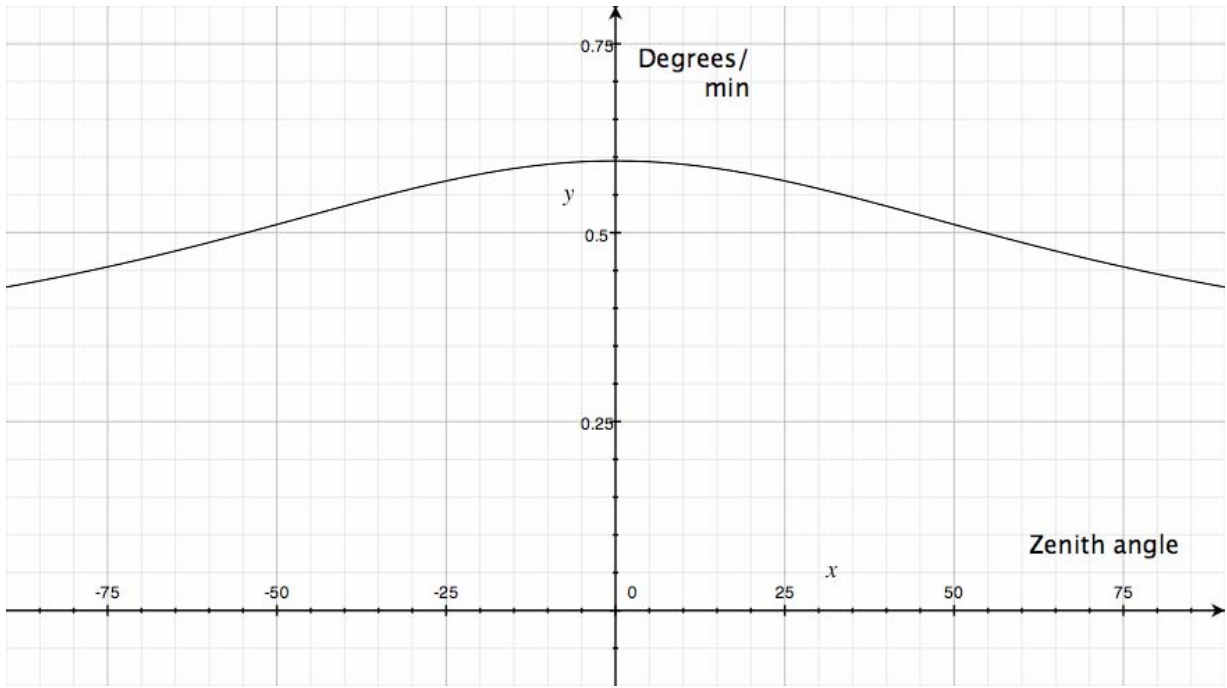

*Figure 6*: Slewing rate as a function of zenith angle for a 20000 km orbit. At the zenith, the satellite's angular velocity with respect to a ground observatory is higher than when it is near the horizon.

Calibration of the SNAP/JDEM space telescope requires an orbit that is at least ~1° away from the Earth's cross-section when viewed from the L2 Lagrange point (1.5 million km from Earth) [16]. This implies an orbit that is at least 20000-30000 km, which a 20000 km orbit would (just barely) suffice. We believe the 1° criterion is very conservative, and thus a 20000 km orbit will indeed allow calibration of SNAP/JDEM in addition to ground-based telescopes, however further studies may be necessary.

Lasers typically have a beam divergence of no smaller than O(1 mrad) without additional optics placed in the beam. However, by expanding the beam, one can reduce this divergence by up to a factor of nearly 1000. Assuming perfect optics, the divergence of a beam is reduced by a factor equal to its expansion; a 1 mm diameter laser beam with 1 mrad divergence, expanded to 50 cm diameter via defocusing optics, will have a divergence of 2 urad [17]. Neglecting atmospheric distortion (discussed in Sec. 7), such a beam would produce a spot of 20 m radius on the Earth directly below the satellite, suitable for calibration of an entire focal plane.

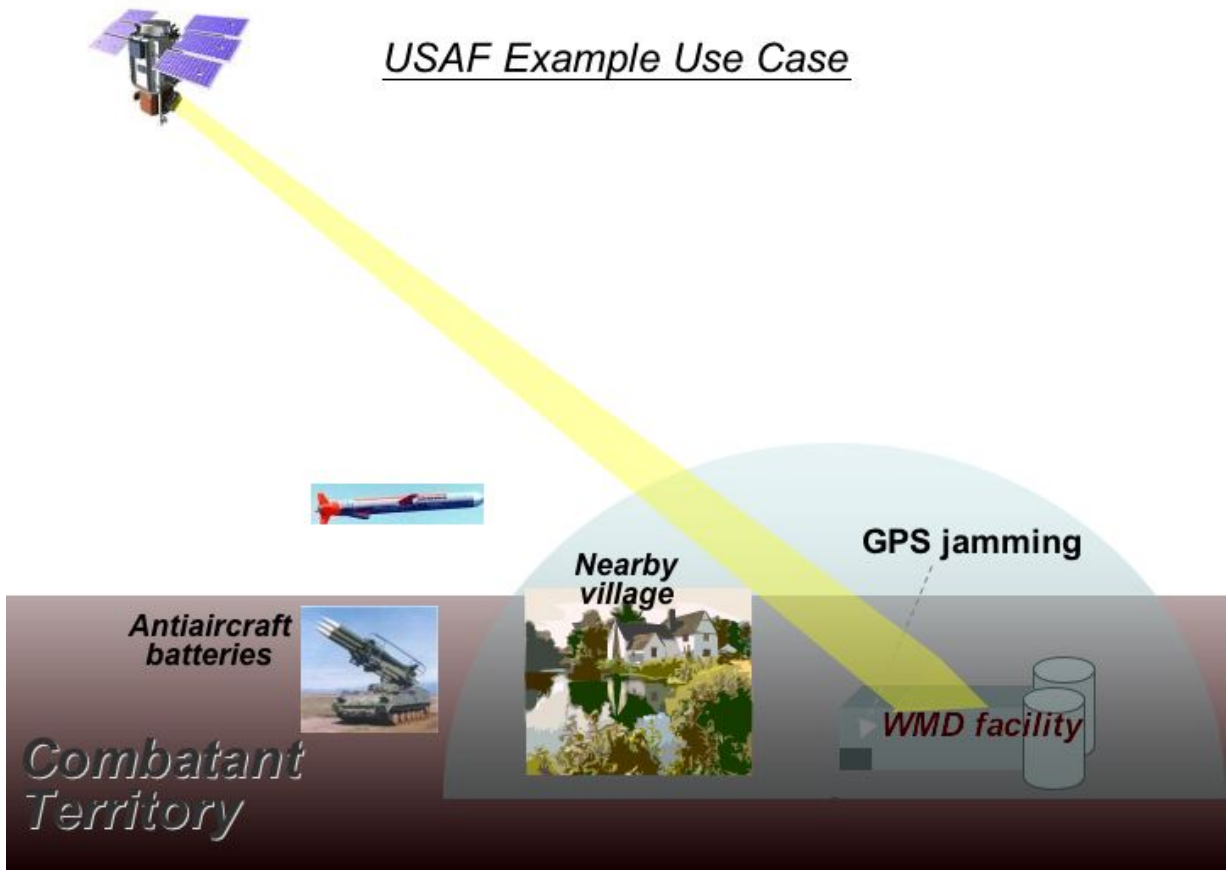

*Figure 7*: Use case example for USAF tactical mission. Forces are ordered to neutralize a nuclear weapons facility (following a major deterioration in diplomatic situation). GPS signals in the vicinity of the facility are intentionally jammed by opposing nation or combatants. Inaccurate targeting could result in collateral damage or even possible accidental radiation spread; thus precise targeting is highly critical. Due to active antiaircraft installations, a manned bomber flight with sight or laser targeting is strongly disfavored. A stand-off cruise missile is preferred, but requires extremely accurate active guidance, despite GPS jam. Onboard TV camera guidance is well known to be unreliable when only satellite photos are available. A laser from space for direct target illumination is the best option, allowing O(20m) remote targeting even in the presence of GPS disruption.

## 3. Military target illumination

In addition to telescope calibration, such a satellite-mounted laser would provide direct tactical benefit to U.S. forces. In realistic scenarios (such as Fig. 7), the presence of GPS guidance jamming on high-value targets dramatically reduces the accuracy and effectiveness of stand-off weapons such as cruise missiles, requiring placement of manned aircraft in harm's way in order to destroy a target. The use of a laser beam from space in place of, or in addition to, satellite guidance would provide an weapon guidance system to ensure precise placement of remotely-launched weapons on target even in the presence of GPS signal disruption. Such a system would undoubtedly save both lives and cost via reducing the amount of otherwise required support for manned bomber missions into enemy territory. Additionally, satellite-guided "smart" bombs from such manned bombing missions would become a great deal smarter due to gaining the ability to accurately strike targets at O(20m) accuracy even in the presence of a GPS signal jam.

## 4. Laser requirements

The tunable laser on a calibration satellite must satisfy two very different goals: the extremely low power and well-calibrated flux in a broad range of wavelengths needed for telescope calibration, and the comparatively higher power needed for target illumination and communication. Both goals also require low dispersion. A typical star or supernova viewed by LSST or Pan-STARRS is approximately 19-23 apparent magnitude, which corresponds to a laser power of only O(10 picowatts) on an 8m ground telescope. In contrast, successful illumination of a ground target in daylight through the atmosphere and above noise sources requires a power of at least O(100 milliwatts).

Clearly the laser itself cannot produce a well-calibrated and consistant power across 10 orders of magnitude. The way to achieve both these very different goals is clearly to have a set of neutral density filters that can attenuate the beam by such a factor, depending on whether they are rotated into the beam or not. A filter wheel (or "filter shutter," as there need only be two different positions) is clearly not a simple device, as it must be extremely carefully designed so as to not affect the satellite momentum. However, this is certainly not new technology, as shutters and filter wheels are in common use on many satellites.

Satellite momentum considerations also necessitate the use of a tunable semiconductor laser rather than a dye laser, as the latter would contain moving liquid dye, whereas the former has no moving parts. Furthermore, the laser must be able to survive, and remain well calibrated, in the high-radiation environment of medium Earth orbit. Tunable semiconductor lasers that produce a stable flux at power of O(100 mW) across a broad range of wavelengths, are indeed available, for example the Light Age 101-PAL tunable semiconductor laser system [17]. Careful testing and studies would be required to ensure the laser maintains flux, dispersion, and pointing characteristics in a long-term high-radiation environment. However, none of the laser components appear at first sight to potentially be highly radiation-sensitive.

## 5. Pointing requirements

To retain stability to O(20m) on the Earth's surface, precise pointing via gyroscopic attitude control must be maintained. From an orbit of 20000 km, 20 m on the Earth's surface corresponds to 1 urad = 0.2 arcsecond stability.

The gyroscopic system on the Hubble Space Telescope achieves stability control to better than 0.005 arcsecond, or 40 times better than this requirement [18]. While the HST uses a star guider to achieve this stability, whereas this calibration satellite must retain pointing stability with the Earth's surface rather than distant stars, stability on this scale and better with respect to Earth's surface is commonly achieved by Earth imaging and reconnaissance satellites (such as the ALOS Earth Imaging Satellite, which achieves stability at 0.07 arcseconds [19]) using fairly similar electromechanical gyroscopic systems to those on the HST, and should not pose significant technology challenge or risk.  Note that O(20m) pointing precision would indeed result in significant movement of the beam with respect to the telescope itself, however the primary purpose of the laser calibration is photometric rather than astrometric (the laser itself does not provide the absolute flux calibration; this is done by a separate calibrated light source, see below), so the O(20m) precision is primarily intended simply to keep enough of the beam on the telescope, as the motion is relatively independent of the observed color.  (Also, multiple measurements and combined fits of the resulting images should allow one some averaging over this astrometric variation.)

## 6. Calibrated diode and LEDs

A NIST-calibrated photodiode [20], to provide an onboard calibration of tunable laser output, is an integral part of ensuring that the flux scale of the laser remains well understood.  In the past decade, the NIST standard for radiometric flux determination has shifted from a blackbody reference to a system that is detector-based; thus the NIST photodiodes are guaranteed to ensure a flux that is properly measured and calibrated to the international standard.  A setup similar to the right-hand side of Fig. 4, with a partially-silvered mirror reflecting a known fraction of the flux back to the calibrated photodiode, can be used.  (The reflected flux fraction will depend on the wavelength, but this function can be calibrated in vacuum on the ground, as the reflectance stability over time of mirror coatings in space is well-known and excellent.)

In addition to the laser-photodiode system, it would undoubtedly be useful to additionally have a calibrated light source aboard the satellite to provide an absolute radiometric flux source for comparison.  For an equivalent brightness to a 23 magnitude star, a simple light source radiating into $4\pi$ solid angle onboard a satellite at 20000 km orbit must have a total output power of 150 watts (while on).  Clearly such an array could only be on for brief periods to avoid draining the satellite power (and turning it off would be an excellent way to subtract off any contributions from satellite albedo).  A light source calibrated to 150 W total output power would undoubtedly be an extremely useful addition to the satellite.  An absolute radiometric flux source for SNAP/JDEM, as opposed to ground-based telescopes, would clearly need to be much brighter; having a $4\pi$ solid angle light source visible at L2 for the time needed to integrate a CCD image would require many times more power than a satellite could provide.  Thus the laser-photodiode setup should be capable of providing a well-calibrated absolute flux by itself.  (If an additional absolute flux source were needed for a cross-check for SNAP/JDEM, a light source integrated within the SNAP/JDEM satellite would likely be the best solution.)

The radiation sensitivity of both NIST-calibrated photodiodes and a calibrated ~150W light source requires study to determine the most radiation-insensitive options.

## 7. Laser divergence and atmospheric dispersion

Two sources of divergence can cause the laser beam to spread before reaching the Earth: intrinsic and atmospheric. Intrinsic divergence of laser light is, as mentioned in Sec. 1, typically O(1 mrad) prior to beam spreading optics, and reduced down to below 2.5 urad by a beam spreader.

In clear conditions, the typical divergence of a laser beam through the Earth's atmosphere is 1 arcsecond [21], or 4.8 urad. So the atmospheric divergence term will dominate over the intrinsic divergence. However, the atmospheric divergence will only act on the second half (approximately) of the laser trip from 3000 km orbit to Earth. Thus, in clear conditions, the spread due to atmospheric divergence will be roughly equal to the intrinsic beam spreading, and maintain the O(20m) beam spot on Earth. This is of course perfectly suitable for flux calibration measurements, as the amount of beam spreading will allow proper calibration of the atmospheric transmission function. For military target illumination, atmospheric divergence will necessitate use of targeting functionality only under relatively clear atmospheric conditions (as is true for any laser guidance system. Note that the choice of wavelength from the tunable laser potentially allows optimization/minimization of the atmospheric dispersion and thus potential guidance use even in somewhat poor atmospheric conditions.)

## 8. Telemetry requirements

Communication with the satellite requires some onboard telemetry. The satellite needs to be able to receive signals to turn the laser on and off, to change frequency, to open and close the neutral density filter "shutter", to turn the LED array on and off, and ground coordinates for an Earth-following slew. It also needs to be able to send signals corresponding to the readout from the calibration photodiode. The latter functionality can potentially be achieved via the laser itself: the laser can encode a pulsed signal corresponding to a readout of the calibrated photodiode without need for separate downlink equipment. The uplink, however, does require some additional equipment. An uplink could either be done via a traditional ground station antenna, or by ground-based laser. The data rate would be not need to be high at all, as the above commands can be encoded within O(10 bits) of data, so an extremely simple uplink receiver should be all that is necessary.

## 9. Possible additional satellite functionality

In addition to calibration of optical and near-IR telescopes, a satellite-mounted laser also has the potential for calibrating cosmic ray telescopes, for example the Pierre Auger Observatory. Cosmic ray fluorescence detectors use lasers fired *upwards* into the atmosphere to calibrate energy scale [22], however the typical systematic uncertainties on energy scale at the upper end of the cosmic ray spectrum are still of order 50% [23]. These could potentially be greatly improved by a satellite-mounted laser firing *downwards* through the atmosphere directly into a fluorescence detector to more directly calibrate the total amount of energy received.

Also, a calibration satellite has the potential to carry other useful, but simple, devices besides a tunable laser, a calibrated photodiode, & LED array. Very simple equipment that may prove useful on such a satellite include a retroreflector panel and a flat optical mirror, for ranging studies and additional atmospheric transmission studies. Additionally, calibration sources for microwave, infrared, and UV regions of the spectrum would potentially be useful, as those regions of the spectrum also suffer from the lack of well-characterized absolute calibration – assuming such sources do not interfere with the operation or lifetime of the optical and near-IR calibration.

## 10. Conclusion

A calibration satellite can fulfill the goals of both the astronomical and military communities, providing the needed absolute flux calibration for precision studies of the fate of the Universe; as well as saving both American and foreign civilian lives by providing a new precision guidance and communications technology. Such a device would help ensure our nation's security, both in national defense, and as the world's science and technology leader.

We hope to submit an expanded version of this white paper as a BAA response or unsolicited proposal to an appropriate program office, and will include an engineering development roadmap as part of that expanded version.

## 11. Acknowledgements

The authors wish to thank Christopher Stubbs (Harvard), David Burke (SLAC/KIPAC), Steven Kahn (Stanford/SLAC/KIPAC), Mandeep Gill (Berkeley/Caltech), John Tonry (Hawaii), Doug Welch (McMaster), Susana Deustua (AAS), Kem Cook (LLNL), Tony Tyson (UC Davis), Nick Kaiser (Hawaii), James Taylor (Caltech), Alan Weinstein (Caltech), and Gregory Dubois-Felsmann (SLAC) for their numerous extremely insightful and helpful comments.

Please note that, since this document was written, the authors have found information on the TSAT satellite project [24], a DoD-supported project for 5 satellites to be launched in 2012-2014 to provide communications replacements for the present MILSTAR satellite network [25]. As these TSAT satellites will communicate via laser links, they might possibly provide an attractive platform for calibration devices such as the above. (This, if possible, would consolidate the above functions into a satellite system already slated for construction, thus potentially saving the cost of a launch.) Further study is needed to investigate this possibility. Additionally, the GPS-3 upgrade to the present GPS satellites [26] will begin launch at a similar time (~2012) and potentially may provide an option with a somewhat lower degree of national security clearance required.